\documentclass{aa}  

\usepackage{graphicx}
\usepackage{txfonts}
\usepackage{amsmath}
\usepackage{color}
\newcommand{\etachi}{\eta_\chi}
\newcommand{\mchi}{m_\chi}
\newcommand{\mprot}{m_p}
\newcommand{\mchizero}{m_{\chi 0}}
\newcommand{\mpzero}{m_{p0}}
\newcommand{\etab}{\eta_b}

\newcommand{\lya}{Ly$\alpha$}

\newcommand{\om}{\Omega_M}

\newcommand{\ocdm}{\Omega_{\chi}}
\newcommand{\ob}{\Omega_B}

\newcommand{\ol}{\Omega_\Lambda}

\newcommand{\rhob}{\rho_{\rm B}}
\newcommand{\rhog}{\rho_\gamma}

\newcommand{\lcdm}{$\Lambda$CDM~}

\newcommand{\mpl}{m_{\rm Pl}}

\begin{document}

\title{Which fundamental constants for  CMB and BAO?}

   \author{
James Rich\inst{1}
          }

\institute{IRFU-SPP, CEA Saclay, 91191 Gif-sur-Yvette, France}

  \abstract{
We use the three-scale framework of Hu et al.
to show how
the cosmic microwave background anisotropy spectrum
depends on the fundamental constants.
As expected, the spectrum 
depends only on
dimensionless combinations of the constants,
and we emphasize the points that make this
generally true for  cosmological observables.
Our analysis suggests that the CMB spectrum shape is mostly determined by
$\alpha^2m_e/m_p$ 
and by $m_p/\mchi$, the proton-CDM-particle mass
ratio.
The distance to the last-scattering surface depends on
$Gm_p\mchi/\hbar c$, so
published CMB observational
limits on time variations of the constants
implicitly assume the time independence of this quantity,
as well as a flat-\lcdm~cosmological model.
On the other hand,
low-redshift BAO, $H_0$, and 
baryon-mass-fraction measurements
can be combined
with the shape of the CMB spectrum to give information
that is largely independent of these assumptions.
In particular, we show that the pre-recombination values of
$G\mchi^2/\hbar c$, $m_p/\mchi$,  and $\alpha^2m_e/m_p$
are equal to their present values at a precision  
of $\sim15\%$.
}


\authorrunning{J. Rich}
\titlerunning{Constants for CMB}

   \maketitle
%
%________________________________________________________________

\section{Introduction}

The cosmic microwave background (CMB) anisotropy spectra are primarily used to
determine cosmological parameters 
\citep{2014A&A...571A..16P,2015arXiv150201589P},
but the spectra can also
give information on the values of the fundamental constants
in the early universe.
Limits on the difference between the pre-recombination and present values
of the fine structure constant, $\alpha$, were first
obtained in studies using CMB data from
BOOMeranG and MAXIMA 
\citep{1999PhRvD..60b3516K,2000PhRvD..62l3508A}
and 
WMAP 
\citep{2004MNRAS.352...20R}. 
The limits were generalized to combined limits on $(\alpha,m_e)$ using
WMAP data 
\citep{2006PhRvD..74b3515I,2008PhLB..669..212S,2009MmSAI..80..814S,2010JCAP...01..030N,2010A&A...517A..62L,2013MNRAS.434.1792S}
and Planck data 
\citep{2015A&A...580A..22P}.
These limits are based on the effects of $(\alpha,m_e)$
on the recombination 
process 
\citep{1999PhRvD..60b3516K,1999PhRvD..60b3515H,2000ApJS..128..407S}.
While the procedure used to obtain these limits is not obviously
incorrect, the publication of a limit on the variation
in $m_e$ is perplexing since it is generally admitted
that only dimensionless fundamental constants are physically
meaningful \citep{1962PhRv..125.2163D}.
This is manifestly true for laboratory measurements, which consist
of comparing quantities of a given dimension with standards
of the same dimension \citep{2003AmJPh..71.1043R}.
It is less obviously true for cosmological measurements where
two times are typically involved.  For example, CMB measurements
concern the time of photon-matter decoupling, $t_{dec}$, and the measurement
time, $t_0$, and one can form dimensionless quantities like
$m_e(t_{dec})/m_e(t_0)$.  In fact, 
CMB-based limits like those of \citet{2015A&A...580A..22P} 
are generally expressed as
limits on the deviation from unity of this dimensionless quantity. 
Similarly, limits from other studies
on time variations of Newton's constant $G$ 
(for a review, see \citet{2011LRR....14....2U}) 
are typically expressed
as measurements of $G(t)/G(t_0)$.
In this paper we  show how a proper analysis gives only
measurements of equal-time dimensionless quantities like $m_e(t)/m_p(t)$.

Part of the problem with using CMB data is that the phenomenology
is rather complicated so it is difficult to include the effects
of all relevant fundamental constants in compact formulas.
This is one reason that results are expressed in terms of dimensioned
constants like $m_e$, since the standard 
numerical procedures like CAMB (http://camb.info) and RECFAST 
\citep{2000ApJS..128..407S} 
use such quantities.
Here, this problem is avoided by using the qualitative 
model of Hu et al. 
\citep{1997Natur.386...37H,1997ApJ...479..568H,2001ApJ...549..669H}
to give the dominant dependencies
of the spectrum on the relevant physical and cosmological parameters.
This  allows us to give a general analysis of the problem, while
the
published studies leading to limits in $(m_e,\alpha)$ space assume the
time independence of all non-electronic masses and of $G$.
Because of these assumptions,
\citet{2015A&A...580A..22P}
interpreted their 
limits on $m_e$  as limits
on $Gm_e^2/\hbar c$, to which one must add the caveat
that all non-electronic masses are held constant.
Quoting limits on $Gm_e^2/\hbar c$ is troubling because gravitational
interactions of electrons should have negligible effects on the
spectrum.
In fact,
the analysis presented here suggests that the
natural dimensionless variables for studying the shape
of the spectrum are  $\alpha^2m_e/m_p$, $m_p/\mchi$ and  
$G\mchi m_p/\hbar c$, where $\mchi$ is the mass of the CDM particles.
The introduction of $\mchi$ into the problem
reminds us that not  even the present values of
all relevant fundamental constants are known. 
However, this
does not prevent us from studying their time variation.

In the following  analysis, Section \ref{parasec} defines the 
fundamental and cosmological parameters, and section \ref{humodelsec}
applies the model of Hu et al.  
to
determine the dependencies of  the CMB spectrum
on those parameters.  
Section \ref{cmbanalysissec} describes the information
that can be derived from an analysis of the spectrum.
Section \ref{limitssec} combines the CMB-derived quantities
with low-redshift measurements to derive limits on the
time variations of fundamental constants.
Finally, Section \ref{conclusion} 
concludes with some thoughts on why cosmological
observations always conspire to give information only on
dimensionless constants.

\section{The fundamental constants and cosmological parameters}
\label{parasec}

We first define the physical and cosmological model that we use.
For the CMB, the five most important coupling constants and masses are
\begin{equation}
\alpha \hspace*{5mm}
G\hspace*{5mm}
\mchi \hspace*{5mm}
m_p \hspace*{5mm}
m_e \; . \hspace*{5mm}
\end{equation}
Since we  allow for time variations,
the  current values are given
with a zero subscript, e.g. $\mpzero$.
Of the five, only $\alpha$ is dimensionless and our goal
is to show that observable quantities depend only on dimensionless
combinations of the last four like $m_e/m_p$ and $G\mchi^2/\hbar c$.
(In this paper, the factors of $\hbar$ and $c$ are generally omitted,
so $G\mchi^2$ is dimensionless.)

As emphasized, for example, in \citet{2011LRR....14....2U},
simply knowing the dependence of observable quantities on fundamental
constants in the absence of time-variations does not mean that one
can reliably calculate the cosmological consequences of time variations.
This is because the physical introduction of time-variations of constants
generally requires the introduction of extra degrees of freedom, like
time-varying scalar fields.  This  adds additional terms to 
the Friedman equation, modifying the expansion rate.
In the absence of a specific model, one has to avoid these complications
by making simplifying assumptions.
As was done in the WMAP and Planck studies 
\citep{2015A&A...580A..22P}
we assume that time variations
of fundamental constants are such that they are effectively time-independent
at high redshift, where they determine the recombination process.
They then quickly ``relax'' to their post-recombination values
where they determine the distance
to the last-scattering surface and provide standards for 
local measurements of the CMB temperature, $T_0$, and the
expansion rate, $H_0$.
We ignore the modifications of the expansion dynamics that
necessarily occur during the relaxation.  This does not significantly
affect our results since we are concerned mostly with
distance-independent constraints.

We assume that the universe at recombination contains baryons, cold-dark
matter particles, photons and neutrinos.
Such a universe is described by 
$\etab$, the
baryon-photon number density ratio, $\etachi$, the same
quantity for dark-matter particles, and $N_\nu$, the number of
neutrino species that were in thermal equilibrium with
the photons for $T>\sim$MeV.
We suppose throughout this paper that $\etab$ and $\etachi$ 
are time-independent.
At least two parameters are necessary to describe the  
primordial fluctuations but these have only a small effect
on our discussion.
The important cosmological parameters are therefore
\begin{equation}
\etachi \hspace*{5mm}
\etab \hspace*{5mm}
N_\nu \hspace*{5mm}
H_0 \hspace*{5mm}
T_0 \hspace*{5mm}
\end{equation}
where $H_0$ and $T_0$ are the current expansion rate and temperature.
The proton and cold-dark-matter masses only enter through 
the gravitational effects of their densities,
$\propto m_p\etab$ and $\propto\mchi\etachi$. 
The most important combinations of physical and cosmological parameters
are therefore $H_0$, $T_0$ and 
\begin{equation}
G\mchi\etachi \hspace*{5mm}
Gm_p\etab \hspace*{5mm}
N_\nu \hspace*{5mm}
\alpha^2m_e 
\end{equation}
where we have anticipated 
that the combination of $(\alpha,m_e)$ that is most
relevant is $\alpha^2m_e$.
We note also that standard studies
replace  $\mchi\etachi$ with 
$\ocdm H_0^2$ by assuming that $G=G_0$:
\begin{equation}
(\ocdm H_0^2)_{no-var} = 2.04 G_0 \mchi\etachi T_0^3 \;,
\label{omh2eq}
\end{equation}
where here and throughout the subscript $no-var$ denotes results
assuming no time variations of fundamental constants.

Because we are mostly concerned with the shape of 
the CMB spectrum, the density of dark energy is
not be an important parameter, since it only enters
into the distance to the last-scattering surface,
determining the angular scale of the spectrum.
However, we  sometimes give results that
depend on this scale, assuming a flat-\lcdm~universe.
In this case, the vacuum energy density 
is  $\ol H_0^2=H_0^2 - \om H_0^2$ where $\om=\ocdm+\ob$.

\section{The CMB anisotropy spectrum}
\label{humodelsec}

To understand the CMB anisotropy spectrum, we use the qualitative model 
of Hu et al 
\citep{1997Natur.386...37H,1997ApJ...479..568H,2001ApJ...549..669H}
based on three
length scales  that are imprinted on the spectrum.
The scales are the Hubble length at matter-radiation equality, $r_{eq}$;
the acoustic scale, $r_A$, equal to the distance a sound wave can travel
before photon-matter decoupling;
and the damping scale, $r_{damp}$, due to photon random walks
near decoupling.
In the anisotropy power spectrum, $C_\ell$,
the three length scales  are reflected in 
three inverse-angular scales,
$\ell_i\sim \pi D(z_{dec})/r_i$, ($i=eq,A,damp$)
where $D(z_{dec})$ is the co-moving angular-diameter
distance to the last-scattering surface.

Besides the three scales, the spectrum depends on four other
parameters: the primordial amplitude of scalar perturbations
and its spectral index $(A_s,n_s)$; 
the effective number of neutrino species, $N_\nu$;
and the baryon-photon ratio at photon-matter decoupling
\begin{equation}
R_{dec} = \frac{3\rhob(T_{dec})}{4\rhog(T_{dec})} = 0.278\frac{m_b\etab}{T_{dec}}
\; .
\label{rdecdef}
\end{equation}
The shape of the spectrum depends on distance-independent
quantities: $r_{eq}/r_A$, $r_{damp}/r_A$, $R_{dec}$,
$N_\nu$ and $n_s$.

Hu et al. propose an approximate  form for $C_\ell$ which
depends on these parameters.
The characteristic peak-trough structure is described by $A_\ell^2$ where
\begin{equation}
A_\ell \propto
[1+R_{dec}T(\ell/\ell_{eq})]\,\cos\pi(\ell/\ell_A + \delta) 
- R_{dec}T(\ell/\ell_{eq}) \;.
\end{equation}
The peaks in the spectrum are at integer values of $\ell/\ell_A+a=n$
where $\delta\sim 0.267$ has only a weak dependence on fundamental and cosmological
parameters.
The cross-term in $A_\ell^2$ favors odd-$n$ (compression) peaks
compared to even-$n$ (rarefaction) peaks  with
the amplitude difference governed by $R_{dec}T(\ell_A/\ell_{eq})$.
Here, $T$ is the matter transfer function expressed in
angular variables, i.e. $T(k/k_{eq})$ with  $k=\ell/D(z_{dec})$.

Averaged over peaks and troughs, 
the amplitude of the spectrum is determined by the other scales,
with $r_{eq}$ governing the rise with $\ell$ above
the low-$\ell$ Sachs-Wolfe plateau and $r_{damp}$ governing the decline
at high $\ell$:
\begin{equation}
C_\ell \propto\ell^{n_s-1} D_\ell^2 P_\ell 
\left[ \frac{A_\ell^2-1}{1 + (\ell_A/2\ell)^6} + 2\right]
\label{huformula}
\end{equation}
where $n_s\sim0.97$ is the spectral index and the ``radiation driving'' and
damping envelopes are
\begin{equation}
P_\ell=\ell^{n_s-1}[1+B\exp(1.4\ell_{eq}/\ell)]
\hspace*{5mm}
D_\ell = \exp[-(\ell/\ell_{damp})^{1.2}]
\label{envelopes}
\end{equation}
where $B\sim 12$ depends on $N_\nu$ and $R_{dec}$ 
\citep{1997ApJ...479..568H}.
Roughly speaking, for $n_s\sim 1$, 
a measurement of the amplitude of the first peak 
relative to the Sachs-Wolfe plateau determines
$\ell_{eq}/\ell_A$  and a measurement of the ratio the higher peaks to the first
determines $\ell_{damp}/\ell_A$.
For models approximating with the observed CMB spectrum,
the
values are $(\ell_{eq},\ell_A,\ell_{dec})\sim(150,300,1300)$ 
\citep{2001ApJ...549..669H}.

We now discuss how the parameters in the expression for $C_\ell$ depend
on the fundamental and cosmological parameters.
The three length scales ($r_{eq}$, $r_{A}$, $r_{damp}$) 
are closely related 
to  the Hubble lengths at
matter-radiation equality, $1/H_{eq}$, 
at baryon-photon equality, $1/H_{p\gamma}$,
and at photon-matter decoupling, $1/H_{dec}$.
They have the simple dependencies on fundamental
and cosmological parameters shown in 
Table 1.
The first column gives the temperatures 
at the redshift where the scales are defined.
The second column gives the inverse scales redshifted to
present epoch where, along with the distance $D(z_{dec})$, they
determine the observed spectrum.
We note the important fact that after this redshift only 
dimensionless combinations of fundamental constants appear
in the second column.

The matter-radiation equality
scale, $r_{eq}$,
determines the minimum $\ell$  that benefited from radiation
driving (early-time Sachs-Wolfe effect), 
resulting an enhancement of the temperature anisotropies
over the primordial value $\Delta T/T\sim10^{-5}$.  
The temperature at equality is
\begin{equation}
T_{eq}=\frac{\mchi\etachi + m_p\etab}{2.7(1+0.68N_\nu/3)} \hspace*{5mm}
\end{equation}
where $N_\nu\sim3$ is the number of neutrino species.
The equality scale is then
\begin{equation}
r_{eq}\equiv \frac{c}{H_{eq}}\frac{T_{eq}}{T_0}= 
\left[ 0.95 \, \frac{\sqrt{G}(\mchi\etachi+m_p\etab)}
{1+0.13\Delta N_\nu} T_0 \right]^{-1}
\label{req}
\end{equation}
where $\Delta N_\nu=N_\nu-3$.

The acoustic scale, $r_A$, is the distance a sound wave can
travel before recombination and determines
the positions of the peaks in the spectrum.
It is determined by two scales: the Hubble scale at
the epoch of baryon-photon equality
(when the sound speed starts to fall below its high-temperature value
of $c_s=c/\sqrt{3}$) and recombination (drag epoch) when the waves stops. 
The first factor is
\begin{equation}
r_{p\gamma} = \left[ 
\sqrt{G(\mchi\etachi+m_p\etab) m_p\etab} T_0 
\right]^{-1}
\label{rpgamma}
\end{equation}
Including the propagation at reduced speed until decoupling gives
\citep{1998ApJ...496..605E}
\footnote{In this paper, we are not  concerned with the small differences
between the acoustic scale $r_A$ and the sound horizon at the drag epoch, $r_d$,
relevant for BAO studies.}
\begin{equation}
r_A = 1.53 r_{p\gamma}F_A(R_{dec},R_{eq})
\label{soundhorizon}
\end{equation} 
where
\begin{equation}
F_A=
\ln \left[
\frac{\sqrt{1+R_{dec}} + \sqrt{R_{dec}+R_{eq}}}
{1+\sqrt{R_{eq}}}
\right]
\end{equation} 
Here, $3\rhob/4\rhog$ at matter-radiation equality is
\begin{equation}
R_{eq}=(3/4)(1+.68N_\nu/3)\frac{m_p\etab}{\mchi\etachi + m_p\etab}
\end{equation}
The value of $R$ at decoupling 
\begin{equation}
R_{dec} = 0.278\frac{m_p}{T_{dec}}\etab = 
0.278 \frac{m_p\etab}{\alpha^2m_e f_{dec}}
\label{bgdeceq}
\end{equation}
where the decoupling temperature has the form $T_{dec}=\alpha^2m_ef_{dec}$
with $f_{dec}$ being a factor that depends weakly on the fundamental
and cosmological parameters and which we now estimate.

\begin{table}
%[htbp] 
%\label{scaletable}
  \begin{tabular}{|l|l|}
    \hline
$T$    & $H(T)\times(T_0/T)$   \\
    \hline
  & \\
$T_{eq}\sim m_\chi\etachi$    &   
$r_{eq}^{-1}$ $\sim\sqrt{G\mchi^2}\etachi\; T_0$ \\
%\alpha^2m_ef_{dec}$  \\
 &  \\
$T_{p\gamma}\sim m_p\etab$ & 
$r_{p\gamma}^{-1}$ 
$\sim\sqrt{G\mchi m_p}\sqrt{\etachi\etab}\;T_0$ \\
%\alpha^2m_ef_{dec}$ \\
 &  \\
$T_{dec}=\alpha^2m_ef_{dec}$ & 
$r_{dec}^{-1}$ $\sim\sqrt{G\mchi m_e\alpha^2f_{dec}}\sqrt{\etachi}\;T_0 $ \\
&  \hspace*{5mm} $\sim r_{p\gamma}^{-1}/\sqrt{R_{dec}}$ \\
 &  \\
\hline
 &  \\
 & 
$D^{-1}$ 
$\sim\sqrt{G_0 \mchizero\mpzero}\sqrt{\etachi T_0/\mpzero}\;T_0$ \\
%$\DA^{-1}$ 
%$\sim\sqrt{G_0 \mchizero\mpzero\etachi T_0/\mpzero}\;\alpha^2m_ef_{dec}$ \\
 &  \\
%$T_0$ & $H_0^2$ &  \\
% & & \\
    \hline
  \end{tabular}
\caption{
Scales relevant for the CMB temperature anisotropy spectrum
}
\tablefoot{
Col. 1: the temperature scale.
Col. 2: the associated distance scale, $1/H(T)$, 
redshifted to the present epoch.
The table shows the
simplified 
dependencies on cosmological and fundamental parameters.
(Numerical factors and factors of $\hbar$ and $c$ are omitted.)  
The redshifting in Col. 2 leaves only dimensionless combinations of 
fundamental constants.
The subscript zero refers to present values and its absence refers to
pre-recombination values.
CDM
domination is assumed $(\mchi\etachi\gg m_p\etab)$.
The factor $f_{dec}\sim0.01$ is a logarithmic function of cosmological
and fundamental parameters, eqn. (\ref{saharesult}).
The fourth line shows the co-moving distance to the last-scattering
surface in the flat-\lcdm~model. 
}
\label{scaletable}
\end{table}

There is no simple approximate formula for $T_{dec}$ because decoupling
happens simultaneously with recombination.  It therefore depends
in a complicated way on the relative rates of recombination, ionization,
and photon scattering.
Simple approximate formulas can be found if one modifies the numerical factors
in the relevant cross sections so that 
one of two extreme conditions is satisfied.
In the first, the 
recombination rates are
sufficiently high to maintain equilibrium abundances of electron and
atoms when decoupling occurs.
In the second, the Compton scattering cross-section is sufficiently
high to decouple the photons after recombination has ``frozen''.
In both cases, one finds that $T_{dec}=\alpha^2m_e f_{dec}$ with
$f_{dec}$ a logarithmic function of physical and cosmological
parameters.

We first consider the case of equilibrium abundances of electrons and
atoms,
so the free-electron density is determined
by the Saha equation.  
The decoupling temperature is defined by equating the photon-electron
(Thompson) scattering rate, $n_e\sigma_Tc$, and the expansion rate.
Using $\sigma_T=(8\pi/3)\alpha^2/m_e^2$ we get
\begin{equation}
f_{dec}^{-1} -3\ln f_{dec} =
2\ln \left[
\frac{8\pi}{3(2\pi)^{3/2}}
\frac{y_e\etab}{\etachi} 
\frac{\alpha^7}{G\mchi m_e}
\right]
\;.
\label{saharesult}
\end{equation}
where $y_e$ is the electron-to-baryon ratio.
For our universe with $m_p\etab\sim\mchi\etachi/5$, this gives
$f_{dec}^{-1}\sim2\ln(\alpha^7/Gm_p^2)\sim 107$.

In the other extreme,
decoupling occurs after recombination reactions
stop.
In this case, one fixes the electron-photon ratio at its value
at ``freeze out'', defined by $H(T_{freeze})=\Gamma(e^-p\rightarrow H)$.
As before with the $T_{dec}$, 
one finds $T_{freeze}=\alpha^2m_ef_{freeze}$ where $f_{freeze}$
is a logarithmic function of physical and cosmological parameters.
The decoupling temperature is then set by diluting the electron
density until $H(T_{dec})=\sigma_Tn_e$ with the result
that $(T_{dec}/T_{freeze})^3=(\langle\sigma v\rangle/\sigma_T)^2$ where 
$\langle\sigma v\rangle$ is
the capture cross-section time velocity at $T_{freeze}$.
As it turns out, the ratio for capture to any bound state is
$(\langle\sigma v\rangle/\sigma_T)^2=\alpha^2m_e/T_{freeze}$ and this
results in $T_{dec}=\alpha^2m_ef_{dec}$
with $f_{dec}=f_{freeze}^{2/3}$ still being
a logarithmic function of physical and cosmological parameters.

In the intermediate, realistic case,
numerical calculations (see e.g. \citet{1999PhRvD..60b3516K})
integrate the Boltzmann equation to find the decoupling temperature.
Studies using Planck and WMAP data use the 
RECFAST code \citep{2000ApJS..128..407S}
which can be modified to include all expected dependencies on the
recombination process on
fundamental constants.
Presumably, such calculations would give a slowly varying dependence
of $f_{dec}$ on fundamental constants as in 
equation (\ref{saharesult}).
The combination  would necessarily be dimensionless and (\ref{saharesult})
suggests that it would be $G\mchi m_e$ times a power of $\alpha$.

The estimate of $T_{dec}$
determines the value of $R_{dec}$ (eqn. \ref{bgdeceq}) and
the damping, $r_{damp}$.
The damping scale is the geometric mean of the photon mean free path
and Hubble scale at decoupling, but at this time the two are forced
to be of the same order of magnitude.
The result is
\begin{equation}
r_{damp} \sim r_{p\gamma}\sqrt{R_{dec}}
\end{equation}

The shape of the CMB spectrum is determined by the distance-independent
ratios of the scales 
in the second column of Table 1, along with $R_{dec}$:
\begin{equation}
R_{dec} =
\frac{m_p\etab}{\alpha^2m_e f_{dec}}
\sim \left(\frac{r_{damp}}{r_{p\gamma}} \right)^2
\label{Rdec}
\end{equation}
\begin{equation}
\frac{r_A}{r_{eq}}=
\left(
\frac{\mchi\etachi + m_p\etab}{m_p\etab}
\right)^{1/2}
F_A(R_{dec},R_{eq})
\label{eqAratio}
\end{equation}
Apart from the weak dependence on $\Delta N_\nu$ and $n_s$, we see that
the spectrum shape is determined by two parameters, 
$\mprot\etab/\mchi\etachi$
and $\alpha^2m_e/m_p\etab$.
Note that $N_\nu$ enters both in the radiation-matter ratio (through $r_{eq}$)
and in the neutrino-photon ratio (through $B$ in equation \ref{envelopes})
so it cannot be absorbed into the other two parameters.

While we are primarily concerned with distance-independent features
in the CMB spectrum, for completeness, we note that
the use of the angular positions of the features induced by these three
scales requires the introduction of the fourth length scale, the distance
to the last-scattering surface.  For flat-\lcdm~models, this is give by
\begin{equation}
D(z_{dec}) = 
\frac{1}{\sqrt{\om H_0^2}}
\int_0^{z_{dec}}\frac{dz}{
\left[
(1-\om)/\om \; +(1+z)^3
\right]^{1/2}
}
\label{d1eq}
\end{equation}
Most of the integral is in the matter dominated redshift range and the
integral is not far from  its value, 1.94, for $\om=1$.
We therefore write
\begin{equation}
D(z_{dec}) = 
\frac{1.94}{\sqrt{\om H_0^2}}
\left[1-f_0(\om)\right]
\end{equation}
where the small correction ranges from $f_0(1)=0$ to $f_0(0.2)=0.13$.

In terms of our adopted cosmological parameters, the distance is given by
\begin{equation}
D(z_{dec})^{-1} = 
\frac{0.82 T_0}{1-f_0}
\left(G_0(\mchizero\etachi+\mpzero\etab)\;  \mpzero \frac{T_0}{m_{p0}}
\right)^{1/2}
\label{d2eq}
\end{equation}
The distance depends on the dimensionless combinations of parameters
$G_0\mchizero\mpzero$
and $G_0\mpzero^2$ and on the measured ratio of the temperature and
the proton mass.

The angular scales associated with the three distance
scales are the ratios between
the length scales and $D(z_{dec})$.
Usually, one refers to the peaks in
$\ell$-space which are near harmonics of $D(z_{dec})/r_A$.
Using (\ref{d2eq}) and (\ref{soundhorizon}) 
we get
\begin{multline}
%\begin{equation}
%\ell_A \sim\frac{\DA}{r_{p\gamma}} = 
\frac{D(z_{dec})}{r_A} \sim
 \left(
\frac{G\mchi m_p}{G_0\mchizero m_{p0}}
\right)^{1/2}
\left(\frac{\etab}
{T_0/m_{p0}}\right)^{1/2}
\frac{1-f_0}{F_A} 
\\
\times \left(
\frac{1 +m_p\etab/\mchi\etachi}
{1 +\mpzero\etab/\mchizero\etachi}
\right)^{1/2}
\label{Doverreqn}
%\end{equation}
\end{multline}
The angular scale thus depends 
on the ratio of $G\mchi m_p$ in the
early universe to the same quantity today.

\section{Analysis of CMB spectra}
\label{cmbanalysissec}

We now reverse the discussion in the previous section and 
discuss the information that can be obtained from the study 
of the observed CMB spectrum.
What one deduces depends on the assumptions made about the 
time-dependence of the fundamental constants and about the
characteristics of  the dark energy.
We consider the three cases:
(1) flat-\lcdm~and no variations of the constants,  
(2) flat-\lcdm~with  variations of
$\alpha$ and $m_e/m_p$ but none of $\mchi$ or $G$, and (3)
all variations allowed and no assumptions on the dark energy or curvature.

The first case corresponds to 
the standard CMB studies that assume no variations and $N_\nu=3$, 
(e.g. \citet{2015arXiv150201589P}).
The CMB spectrum shape can be fit to determine
$\mprot\etab/\mchi\etachi$ and $\alpha^2m_e/m_p\etab$.
Imposing the low-redshift value of $\alpha$, $m_e$ and $m_p$
then determines $\etab$ and $\mchi\etachi$.
Then assuming no evolution of $\mchi\etachi$ and using $G=G_0$
one determines
$\ob H_0^2\propto m_p\etab$ and $\Omega_\chi H_0^2\propto \mchi\etachi$.
This is consistent the well-known fact that the CMB shape determines
precisely these two cosmological parameters, if one assumes that
the fundamental constants have not varied.
That they are determined only by the shape is attested by 
the fact that fits allowing curvature do not change significantly
the central values or errors on $\ob h^2$, $\om h^2$ or $r_A$ 
\citep{2014A&A...571A..16P}
Allowing curvature would permit compensating changes in $D(z_{dec})$
and $r_A$ so as to maintain the angular scale, but this is not
seen because it is the shape that determines $(\ob h^2,\ocdm h^2)$ and, 
hence, $r_A$. 
We note, however, that not requiring $N_\nu=3$ 
increases $\ocdm h^2$ by $\sim5\%$ and doubles its error.
These changes, and the corresponding changes in $r_A$ are sufficiently
small to ignore for
the limits we  find in section \ref{limitssec}.

The second case corresponds to 
the traditional studies of time variations, e.g. \citet{2015A&A...580A..22P},
where one does not impose the local values of $\alpha$ or $m_e/m_p$.
In this case,
the shape-determined values of $m_p\etab/\mchi\etachi$ and 
$\alpha^2m_e/m_p\etab$ are not sufficient to separately measure
the cosmological and fundamental parameters.
These studies therefore also 
use the angular scale, assuming that it is given by
the flat-\lcdm~result (\ref{Doverreqn}) and assume
that $G\mchi m_p$ has not varied in time.
In this case, 
equation (\ref{Doverreqn}) provides a third constraint, determining $\etab$.
The shape-determined value of $\alpha^2m_e/m_p\etab$ 
then determines $\alpha^2m_e/m_p$.
This pre-recombination value
can then be compared with
the $(\alpha^2m_e/m_p)_0$.  This is a simplified
version of what is done
in traditional CMB studies of time variations. 
Studies using  WMAP data 
\citep{2006PhRvD..74b3515I,2008PhLB..669..212S,2009MmSAI..80..814S,2010JCAP...01..030N,2010A&A...517A..62L,2013MNRAS.434.1792S}
confirm that in the $(\alpha,m_e)$
space, the best determined combination is indeed $\sim\alpha^2m_e$.
(Those studies assume a fixed $m_p$.).
The Planck data extends to sufficiently high $\ell$  
to give tight constraints
on other combinations of $(\alpha,m_e)$ \citep{2015A&A...580A..22P}.

We now turn to the last case,
what can be learned if one makes no assumptions
about the time variations of the fundamental constants or the dark energy.
Lacking a consistent analysis of the CMB spectrum leaving all
constants free, we must look for scaling relations that say
how the announced results would be modified if variations are allowed.
Equation \ref{Rdec} suggests that the 
CMB measurement\footnote{We use throughout the ``TT+lowP'' 
values from \citet{2015arXiv150201589P}.  
}
of $m_p\etab$ ($\propto \ob h^2=0.02222\pm0.00023$) 
comes from the baryon-photon ratio $R_{dec}$ and should therefore be
understood as a measurement of  $m_p\etab/\alpha^2m_e$, if we ignore
the weak parameter dependence of $f_{dec}$.
We can interpret the CMB measurement as
\begin{equation}
( m_p\etab)_{no-var} = m_p\etab 
\frac{(\alpha^2m_e)_0}{\alpha^2m_e}
\end{equation}
where the subscript $no-var$ refers to values reported
assuming no time variations.
This formula should be regarded as a first-order approximation, since
we neglect the logarithmic dependence of $f_{dec}$ on the parameters.
CMB studies convert $m_p\etab$ to $\ob H_0^2$ using the
laboratory value of Newton's constant:
\begin{multline}
(\ob h^2)_{no-var} =\frac{2.04 T_0^3 G_0 m_p\etab}
{(100{\rm km\,s^{-1}Mpc^{-1}})^2}
\frac{(\alpha^2m_e)_0}{\alpha^2m_e}
\\
=
0.02222\pm0.00023
\hspace*{30mm}
\label{obplanck}
\end{multline}
where $h=H_0/100{\rm km\,s^{-1}Mpc^{-1}}$.
The baryon mass fraction measured with the CMB spectrum does
not use the value of the proton mass measured at low redshift so
\begin{equation}
\left( \frac{\ob h^2}{\ocdm h^2 } \right)_{no-var} =
\frac{m_p\etab}{\mchi\etachi}
= 0.1856 \pm 0.004  
\label{obomplanck}
\end{equation}
This implies with (\ref{obplanck})
\begin{multline}
%\begin{equation}
%(\ocdm H_0^2)_{no-var} =2.04 T_0^3 G_0 \mchi\etachi
(\ocdm h^2)_{no-var} =\frac{2.04 T_0^3 G_0 \mchi\etachi}
{(100{\rm km\,s^{-1}Mpc^{-1}})^2}
\frac{(\alpha^2m_e)_0}{\alpha^2m_e}
\\
=0.1197\pm0.0022
\hspace*{30mm}
\label{ocdmplanck}
\end{multline}
Finally,
expressing $r_A$ in (\ref{soundhorizon}) in terms of the
directly measured quantities $\alpha^2m_e/m_p\etab$ and
$m_p\etab/\mchi\etachi$, one finds
\begin{equation}
(r_A)_{no-var}=r_A \left( 
\frac{(Gm_e^2\alpha^4)_0}{Gm_e^2\alpha^4}\right)^{1/2}
= (147.33\pm0.49){\rm Mpc}
\label{raplanck}
\end{equation}
Relations (\ref{obomplanck}), (\ref{ocdmplanck}) and
(\ref{raplanck}) are used in the  next section to set limits
on time variations of the fundamental constants.

\section{Limits on time variations}
\label{limitssec}

The CMB derived values in the expressions
(\ref{obomplanck}), (\ref{ocdmplanck}) and
(\ref{raplanck}) can be compared with measurements of the analogous
quantities at low redshift to set limits on time variations of the fundamental
constants that appear in the expressions.  
The fact that measurements of cosmological parameters generally agree
with the ``concordance \lcdm~model'' at the 10\% level 
tells us to expect constraints at this level.
All of these limits  use the locally measured value
of the Hubble constant: 
$H_0=(72\pm3){\rm km\,s^{-1}Mpc^{-1}}$ \citep{2013ApJ...775...13H}.

The most direct limit comes from comparing (\ref{obomplanck})
with the same quantity  derived from the baryon mass-fraction in galaxy
clusters. \citet{2014MNRAS.440.2077M} found 
$h^{3/2}\ob/\om= 0.089\pm0.012$, implying $\ob/\om=0.145\pm0.02$ and
\begin{equation}
\frac{\ob}{\ocdm}\equiv \frac{\mpzero\etab}{\mchizero\etachi}=0.170\pm0.023
\label{obocdmlowz}
\end{equation}
This measurement assumes that galaxy clusters are sufficiently large
to contain a representative sample of all massive species, an assumption
justified by simulations of structure formation.
Dividing (\ref{obomplanck}) by (\ref{obocdmlowz}) and assuming that $\etab$ and
$\etachi$ are time independent gives
\begin{equation}
\frac{m_p/\mchi}{\mpzero/\mchizero} = 1.09\pm 0.15
\label{mpmchilimit}
\end{equation}
While we do not know the value of $\mchi$, this shows that
it is stable in time, relative to the proton mass.
We note however, that there is a controversy concerning
cluster masses \citep{2015arXiv150201024S} so this result should
be considered as provisional.

The use of equation  (\ref{ocdmplanck}) is delicate because there are no
direct low-redshift measurements of the matter density as there
are of the photon density.  The simplest constraints come from
Hubble diagrams using type Ia supernovae or the baryon-acoustic-oscillation
(BAO) standard ruler.  These measurements of the matter density are, of
course, complicated by the fact that dark-energy dominates
at low redshift so the deceleration expected from matter turns out
to be an acceleration..  
It is necessary to make some simplifying assumptions about the dark energy
and we make the usual assumption that it is sufficiently well
described by a cosmological constant, though we make no assumptions
about the curvature, i.e. we do not require $\om+\ol=1$.

The most useful measurements for our purpose is the BAO Hubble diagram
unconstrained by the CMB calibration of $r_A$.
The physics that leads to the peaks in the CMB spectrum also 
generates the BAO peak seen in the correlation
function of tracers of the matter density.  While the non-linear
processes leading to structure formation make the correlation
function more complicated to interpret than the CMB spectrum,
the position of the BAO peak is believed to be placed reliably
at $r_A$ to a precision of better than 1\%.  Unlike the CMB spectrum
which is only observed in the transverse (angular) direction, the
BAO feature can be observed in both the transverse and radial (redshift)
directions.  The observable peaks in (redshift,angle) space 
in the correlation function at redshift $z$ are at
\begin{equation}
\Delta \theta_{BAO}=\frac{r_A}{D(z)} 
\hspace*{5mm}
\Delta z_{BAO}=\frac{r_A}{c/H(z)}
\end{equation}
where we ignore the small difference between $r_A$ and $r_d$, the sound horizon at the drag epoch (slightly
after photon decoupling).
If averaged over all directions (longitudinal and transverse), the BAO
peak measures $r_A/D_V(z)$ where $D_V(z)^3\equiv (z/H(z))D(z)^2$.
% and $D_M(z)=(1+z)D_A(z)$.

Using the available measurements of $D(z)/r_A$ and $c/H(z)/r_A$
one can fit for the two density parameters $(\om,\ol)$
and the sound horizon relative to the present Hubble scale 
$(c/H_0)/r_A)$. The results 
(figure 3 of \citet{2014arXiv1411.1074A}) is
\begin{equation}
\om=0.29\pm0.05 \hspace*{10mm} \frac{c/H_0}{r_A}=29\pm1
\end{equation}
We note that the sensitivity for $\om$ is enhanced by the
measurement of $c/H(z=2.34)/r_A=9.18\pm0.28$ by
\citet{2015A&A...574A..59D} at a redshift
where the universe is expected to be matter dominated.  
The precise measurement of $c/H_0r_A$ is driven by the
$r_A/D_V(z=0.106)=0.336\pm0.015$ from \citep{2011MNRAS.416.3017B} at a
redshift where all distances are to good approximation 
proportional to $c/H_0$.

Using $H_0=(72\pm3){\rm km\,s^{-1}Mpc^{-1}}$ \citep{2013ApJ...775...13H}
gives
\begin{equation}
\om h^2 =  0.150\pm0.026
\hspace*{10mm}
r_A=143.5\pm5.9
\label{lowzrd}
\end{equation}
Removing the baryonic component from $\om h^2$ gives 
$\ocdm h^2=0.128\pm 0.021\propto G_0\etachi\mchizero$.
Comparing this value with  the 
Planck result (\ref{ocdmplanck}) gives
\begin{equation}
\frac{(\ocdm H_0^2)_{no-var}}{(\ocdm H_0^2)_{low-z}} = 
\frac{(\alpha^2m_e/\mchi)_0}{\alpha^2m_e/\mchi}
= 0.93 \pm 0.16
\label{asqmemchilimit}
\end{equation}

Finally, comparing the CMB calculated sound horizon
(\ref{raplanck}) with the low-redshift value (\ref{lowzrd}), we get
\begin{equation}
\frac{r_A}{(r_A)_{no-var}}=
 \left( \frac{Gm_e^2\alpha^4}{(Gm_e^2\alpha^4)_0}\right)^{1/2}
= 0.97\pm 0.04
\label{Gmesqa4limit}
\end{equation}

The three limits (\ref{mpmchilimit}), (\ref{asqmemchilimit}), and
(\ref{Gmesqa4limit}) exhaust the information that we can obtain from
the three-scale model.
For example, we could derive a limit analogous to (\ref{Gmesqa4limit})
with $r_{eq}$ instead of $r_A$ using a the position of $r_{eq}$ in the
matter power spectrum at low redshift 
\citep{2007MNRAS.378..852P,2007MNRAS.374.1527B}.
However, this would not give an independent limit since we have
already used the ratio $r_{eq}/r_A$ in the other limits.

The three limits can be combined to limit time variations on other interesting
combinations, like $G\mchi^2$ and $Gm_p^2$.
In fact, the limits can be summarized as excluding large variations
of all ratios of the four mass scales that enter the problem:
\begin{equation}
\frac{m_i/m_j}{(m_i/m_j)_0} \sim 1.0\pm\sim0.15
\hspace*{5mm}m_i,m_j = m_{pl},\;\mchi,\;m_p,\;\alpha^2m_e
\label{limitsummary}
\end{equation}
where the Planck mass is $m_{pl}=\sqrt{\hbar c/G}$.
The 15\% precision on these limits is dominated by the precision of the
low-redshift measurements and relatively insensitive to small modifications
of the pre-recombination physics.  For example, not requiring $N_\nu=3$ 
increases the uncertainty in the CMB-derived CDM density to $\sim5\%$, still
small compared to the low-redshift uncertainties..

Our limits assume that there are no large changes in the fundamental
constants during late times that would invalidate the interpretation
of the low-redshift measurements.
They could therefore  
%(\ref{mpmchilimit}), (\ref{asqmemchilimit}), and (\ref{Gmesqa4limit}), 
be evaded if the late-time variations
somehow canceled the pre-recombination variations. 
All three  limits use distance-ladder measurements of $H_0$ and the
use of this ladder assumes no variations
of the electromagnetic or gravitational interactions of ordinary
matter, which would affect the luminosities of Cepheid variable stars
and supernovae. 
There are strict limits on variations of such interactions 
at the level of $10^{-12}{\rm yr^{-1}}$
for gravitational interactions \citep{2004PhRvL..93z1101W}
and $10^{-16}{\rm yr^{-1}}$ for
electromagnetic interactions \citep{2011LRR....14....2U}.  
These are stronger that those presented
here which are of order $10^{-11}{\rm yr^{-1}}$. This suggests
that the limit (\ref{Gmesqa4limit}),
which uses only the distance ladder,
is insensitive to our assumption
of no low-redshift variations.  
On the other hand, the two other
limits use the gravitational interaction of dark-matter particles
in galaxy clusters and in cosmological deceleration.  As such, one
cannot appeal to strong limits on current variations
to argue against compensating variations.
Most conservatively, the limits
(\ref{mpmchilimit}), (\ref{asqmemchilimit}), and (\ref{Gmesqa4limit}), 
should then be interpreted as constraints on theories that 
predict both early- and late-time variations.

\section{Conclusion}
\label{conclusion}

The prime motivation of this study was to clear up the question of
what fundamental constants determine the CMB anisotropy spectrum
and to show that they consist of dimensionless combinations.
In this context,
the striking result of this study is seen in the second
column of Table 1:  all three length scales of the CMB spectrum, after
redshifting to the present epoch, depend on dimensionless combinations
of constants in the pre-recombination universe.  Before the redshifting,
the dimensionality was contained in the fundamental constants.
The redshifting transferred the inverse-length dimension to $T_0$.
This means that even if the distance to the last-scattering surface
were somehow known, the angular features would depend only on dimensionless
combinations in the pre-recombination universe.

In fact,
the distance to the last scattering surface must be
calculated. For the flat-\lcdm~model, it is shown in the fourth line of
Table 1.   It also depends on a dimensionless combinations, this time
at the present epoch.
This came about by the ``trick'' of writing
$G\mchizero T_0$ as $G\mchizero\mpzero\times T_0/\mpzero$.
This just corresponds to our freedom  to express
measured quantities like $T_0$ as multiples of fundamental quantities.
In fact, this ``freedom'' is an obligation since it takes into
account the dependence of our SI standards on fundamental constants.
Expressing results in such manifestly dimensionless forms avoids all
discussion about what units are being used. 

The transfer of the inverse-length dimension to $T_0$ works for any
standard ruler, so our conclusion that only dimensionless combinations
are relevant for length scales is quite general.
A similar reasoning works for standard candles \citep{2013arXiv1304.0577R}.
For example, if one can express the total energy output of 
a supernova, $Q_{SN}$, in terms of  fundamental constants
(e.g. $Q_{SN}\sim(\mpl/m_p)^3Q_{56}$, where $Q_{56}$ is the
energy liberated in the $\beta$-decay of $^{56}$Co), then
one can also work with the dimensionless energy output, $Q_{SN}/\alpha^2m_e$.
This quantity gives the number of photons that would be produced if
all energy were converted to \lya~photons.  It can be related
to the true number of photons by scaling by the 
observed ratio of the
mean supernova photon energy to the energy of \lya~photons
from the same redshift.
Therefore, the supernova photon output depends
only on the dimensionless combination $Q_{SN}/\alpha^2m_e$ 
and a directly measurable energy ratio.

The CMB observables studied here are the distance independent quantities
(\ref{Rdec}) and (\ref{eqAratio}) which 
provide a tidy way of summarizing the 
first-order cosmological
and physical information contained in the CMB spectrum.
The combinations of parameters seen in these  expressions 
reflect the degeneracies between fundamental and cosmological
parameters 
that can be broken by explicitly assuming a flat-\lcdm, constant-$G$ model
\citep{2015A&A...580A..22P}.
Here, we have shown how
combinations of CMB data with low-redshift measurements of cosmological 
parameters lead to the more model-independent limits summarized by
(\ref{limitsummary}).
It will be a challenge to incorporate these qualitative results 
into a rigorous analysis
of the CMB spectrum.
Such an analysis would certainly modify two of the scaling relations
we have used, (\ref{ocdmplanck}) and (\ref{raplanck}), because of the
complications in the $\alpha$ dependence of recombination that
we have not taken into account.  This would modify the effective
dimensionless combination of constants that are probed so
the limits  (\ref{asqmemchilimit}) and (\ref{Gmesqa4limit}) should
be viewed as first order results.
The limit (\ref{mpmchilimit}) is more robust cosmologically because
baryons and cold-dark-matter enter the system only through their
densities.  In this case, the limit is accurate only to the
extent that the interpretation of the low-redshift data is reliable.

\begin{acknowledgements}

I thank Nicolas Busca, Sylvia Galli,
Jean-Christophe Hamilton,  Claudia Sc\'occola,
Douglas Scott, and especially 
Jean-Philippe Uzan for helpful comments and suggestions.

\end{acknowledgements}

\bibliographystyle{aa}
\bibliography{constantscmb}

\end{document}